# Structural Exposure and Subsequent Neighborhood Stability in Temporal Networks: Evidence and Limits of Generalizability

Mehmet ŞİMŞEK
ORCID:0000-0002-9797-5028
Sinop University, Faculty of Engineering and Architecture, Department of Computer Engineering,
57000, Sinop, TÜRKİYE

**Abstract**

Can temporal-path-based structural exposure serve as a reliable indicator of subsequent local neighborhood stability across different temporal networks? We investigate this question using the completed observational results from 12 networks spanning face-to-face interaction, communication, online activity, and urban mobility. Exposure aggregates source-node structural mass along time-respecting paths; the outcome is Jaccard neighborhood turnover between later, nonoverlapping windows, with greater turnover indicating lower stability. Six complementary analyses examine the direction and consistency of exposure–turnover associations, proximity to high-mass nodes, functional response shapes, distance-decay kernels, alternative mass definitions, and temporally ordered changes in source mass. The evidence is heterogeneous rather than uniformly negative or positive. In the primary association analysis, two networks retained positive coefficients after within-question false-discovery-rate adjustment, indicating lower subsequent stability at higher exposure, whereas no negative association survived that adjustment. The pooled proximity coefficient was not distinguishable from zero, and significant mass-change associations appeared in opposite directions across networks. More flexible functional, kernel, and mass choices yielded only modest cross-validated improvements over simpler comparators; available temporal randomizations also gave model-dependent contrasts. The study therefore does not establish structural exposure as a directionally consistent, network-independent stability indicator. It also cannot establish a lack of predictive value relative to a no-exposure baseline, because that specific comparison and independent cross-network validation were unavailable. Incomplete null simulations, unrecorded covariance-estimation fallbacks, and the absence of external replication constrain inference. The contribution is an explicitly bounded assessment of when a temporal structural-exposure statistic does—and does not—provide consistent evidence about later neighborhood stability.

Keywords: Temporal networks; Structural exposure; Neighborhood stability; Neighborhood turnover; Temporal paths; Cross-network heterogeneity; Generalizability.

## 1. Introduction

Temporal interaction networks evolve through timestamped events rather than continuously active edges. Face-to-face encounters, emails, online contacts, and movements between locations consequently generate changing sets of neighbors at each node. A practical question is whether structural information recorded in one period can indicate how persistent a node's immediate neighborhood will be in subsequent periods. Answering it requires preserving the chronology of contacts instead of relying solely on a static aggregate, as established in foundational temporal-network research (Holme and Saramäki, 2012). Here, "local structural stability" means persistence of neighbor identities across later observation windows, not invariance of the entire network.

Event timing changes which connections are usable. A static route may not be traversable in the order required by its constituent interactions, and temporally valid routes can have durations unrelated to their static lengths. Pan and Saramäki (2011) documented these differences in communication and transportation networks. Temporal correlations and burstiness have also been shown to shape spreading rates (Karsai et al., 2011). Although these findings motivate temporally informed structural measures, spreading speed and future neighborhood persistence are different outcomes. Their relationships with structural exposure must therefore be tested rather than assumed.

The time terminology in this study is strictly operational. Global time is the timestamp axis of the event stream, partitioned at a dataset-specific resolution. Local turnover measures how much the labeled set of a node's neighbors changes between two subsequent windows; a larger turnover implies lower neighborhood stability. A change rate or relative structural-clock rate may summarize that activity at a fixed resolution, but neither represents physical proper time. The present article tests an indicator of subsequent neighborhood stability; it does not assert an alteration in the flow of physical time.

Our candidate indicator, structural exposure, aggregates the contributions of other nodes according to their structural mass and their reachability along time-respecting paths. The reference construction weights degree-based mass by an inverse-square kernel of normalized temporal duration. Mass and distance are algorithmic properties of the event network, not physical quantities. Importantly, combining temporal distance and node mass is not in itself novel: Bi et al. (2021) proposed a temporal gravity model to identify influential nodes using analogous ingredients. That study evaluated node importance and spreading influence; our outcome instead concerns the identities of a target node's future neighbors. The contribution claimed here is the empirical scrutiny of this particular exposure–stability relationship, not the invention of mass–distance modeling.

Structural exposure might be associated with subsequent persistence for several competing reasons. Access to prominent and temporally reachable nodes may coincide with recurrent contacts and thus greater stability; alternatively, highly connected settings may provide opportunities for partner replacement and thus lower stability. Activity-driven models show why time-varying participation cannot be inferred from aggregate connectivity alone (Perra et al., 2012). Higher-order dependencies and path histories can also affect the adequacy of simple network descriptions (Scholtes, 2017); (Lambiotte et al., 2019). Neither line of research determines the sign of the exposure–turnover relation examined here.

Prior evidence also cautions against expecting one transferable structure–time pattern. Ceria et al. (2022)examined the proximity of contacts in time and topology across 13 empirical temporal networks and reported domain-dependent relationships. That work analyzes correlations among contacts, not a path-weighted indicator of subsequent node-neighborhood retention, so it is related context rather than an identical test. Likewise, investigations of temporal network prediction based on memory (Zou et al., 2023) concern predictive structure under different modeling targets. Taken together, these studies motivate an explicit distinction between an indicator that is associated with an outcome in some networks and one whose direction or predictive benefit is reliable across networks.

For clarity, we use "reliable indicator" as an evaluative question, not as a formally certified property. Evidence in favor would require a reasonably consistent direction of the exposure–turnover association, resilience to sensible changes in the mass and distance definitions, and predictive improvement that persists outside the fitting context. The present archived analyses address the first two issues through network-specific regressions and within-network blocked comparisons, but do not include an external replication result or a direct no-exposure versus exposure predictive benchmark. Consequently, this study can challenge a claim of general directional consistency; it cannot comprehensively establish clinical-style calibration, causal validity, or out-of-network predictive utility.

Null models provide a complementary diagnostic. Timestamp shuffling, rearranging event intervals, or rewiring partner relations disrupt different properties of the same event stream. Their interpretation depends on exactly which properties remain constrained, as detailed by Gauvin et al. (2022). A substantive complication arises when randomization changes which node–window observations meet the analysis eligibility rule: the resulting reference distribution may no longer describe the identical empirical panel. We therefore interpret the available null comparisons as procedure-specific sensitivity evidence, rather than universal tests of indicator validity.

We examine these questions using the available completed analyses for 12 development networks from four interaction domains. In the design, structural exposure is calculated from a historical window and subsequent neighborhood turnover is measured between two later, nonoverlapping windows. The original prespecified directional conjecture predicted that greater exposure would accompany lower turnover, or higher stability. Rather than redefining success after observing the results, we retain that directional benchmark and report coefficients in their originally modeled turnover scale. Under this convention, a positive coefficient corresponds to lower subsequent stability, not greater stability.

Six research questions jointly assess the candidate indicator: whether exposure is associated with later turnover and in which direction (RQ1); whether proximity to high-mass sources exhibits a consistent gradient (RQ2); whether adjusted response curves favor a shared shape (RQ3); whether temporal-distance kernel choice materially improves raw-turnover prediction (RQ4); whether alternative definitions of source mass improve it (RQ5); and whether mass-induced exposure changes precede consistent subsequent turnover changes (RQ6). The last analysis is an observational mass-change decomposition, not a causal perturbation. The completed temporal randomizations are supplementary rather than comprehensive confirmatory tests.

The study offers a bounded empirical evaluation of a specific, reproducible candidate indicator. Its contribution is threefold: it compares the direction and magnitude of exposure–turnover relations across diverse real-world networks; it examines the sensitivity of model fit to exposure construction rather than treating a favored specification as self-evidently transferable; and it documents null-model and computational limitations without treating unfinished runs as negative evidence. Results from the planned independent replication panel and synthetic recovery experiments were unavailable. Our conclusions therefore concern the completed development panel and the modeling definitions used here, not all possible temporal networks or stability indicators.

## 2. Materials and Methods

### 2.1. Study design and operational notion of indicator reliability

This observational study examined whether mass-weighted exposure accumulated along time-respecting paths is a directionally consistent and model-robust indicator of subsequent local neighborhood stability across empirical temporal networks. Stability was interpreted through the inverse meaning of the already defined Jaccard turnover outcome: lower turnover represents more persistent neighbor identities. All archived regressions and predictive comparisons retain turnover as the response; no new stability equation or re-estimated endpoint was introduced. The study assessed network-specific exposure–turnover associations (RQ1), high-mass proximity (RQ2), functional response shapes (RQ3), distance kernels (RQ4), structural-mass definitions (RQ5), and temporally ordered mass-induced exposure change (RQ6). We use the term “indicator” descriptively; this design neither establishes a causal predictor nor supplies independent out-of-network predictive validation. Temporal reference networks were used as partially completed sensitivity analyses.

The configured development panel contained 12 temporal networks spanning face-to-face contacts, electronic communication, online interactions, and urban mobility (Table 1). Directionality followed each data source’s interaction semantics. In the configuration, four additional networks (Copenhagen Networks Study, Enron E-mail, Reddit Hyperlinks, and TfL Santander Cycles) were reserved for a separate replication panel. No numerical estimates from this reserved panel were available in the supplied results archive; consequently, the present manuscript reports only analyses with archived development-panel outputs. All dataset names, directedness flags, and base temporal resolutions in Table 1 correspond to the supplied dataset configuration.

**Table 1. Development networks and configured base temporal resolution.**

| Domain | Network | Directed | Base window, Δ (s) |
|---|---|---|---|
| Face-to-face contact | High School 2013 | No | 600 |
| Face-to-face contact | Workplace 2015 | No | 1,800 |
| Face-to-face contact | SFHH Conference | No | 300 |
| Communication | Email-EU-core | Yes | 21,600 |
| Communication | CollegeMsg | Yes | 21,600 |
| Communication | Manufacturing E-mail | Yes | 86,400 |
| Online interaction | Facebook Wall Posts | Yes | 21,600 |
| Online interaction | MathOverflow | Yes | 43,200 |
| Online interaction | Bitcoin OTC | Yes | 1,209,600 |
| Urban mobility | NYC Yellow Taxi | Yes | 900 |
| Urban mobility | NYC Citi Bike | Yes | 900 |
| Urban mobility | Chicago Divvy | Yes | 900 |

### 2.2. Event processing, observation windows, and eligibility

The dataset-specific importers converted source records to timestamped node-pair events. Self-loops were removed; node pairs in undirected networks were represented canonically; and dataset-specific filters were applied before constructing temporal paths. In particular, the NYC Yellow Taxi importer restricted pickup and drop-off zone identifiers to the configured valid range of 1–263. The three mobility datasets were restricted to the configured April–May 2019 analysis period. Signed or natively weighted event information was retained by the importers where available, although the reference exposure used unweighted degree-based mass. Timestamps were normalized to integer epoch seconds for analysis.

For a network with base window width $\Delta$ and temporal anchor $t$, the primary panel used one historical exposure window followed by two nonoverlapping outcome windows. The exposure window $W_0(t)$ ran from $t - \Delta$ (inclusive) to $t$ (exclusive). The first outcome window $W_1(t)$ ran from $t$ to $t + \Delta$, and the second outcome window $W_2(t)$ ran from $t + \Delta$ to $t + 2\Delta$, in each case excluding the upper endpoint.

An eligible node had at least one incident event in each of the three windows; this three-window activity requirement was reapplied to the randomized networks. Anchors advanced by one base window. Configuration-level adequacy thresholds required at least 100 network nodes, 10,000 events, 50 informative anchors, and 30 nodes common to all three windows at an informative anchor. These were inclusion and computation rules, not guarantees that a coefficient could be estimated for every network and research question. Exposure predictors were calculated from $W_0$, whereas the primary outcome compared $W_1$ and $W_2$; this temporal separation reduces direct overlap between predictor and outcome construction but does not exclude unmeasured confounding.

### 2.3. Local structural turnover and complementary outcomes

Let $N_i^{(1)}(t)$ and $N_i^{(2)}(t)$ denote the sets of distinct labeled neighbors of eligible node $i$ in $W_1$ and $W_2$, respectively. In directed networks, the labels distinguish incoming from outgoing relations; in undirected networks, no direction label is used. The primary outcome, Jaccard structural turnover, was

$$D_i(t) = 1 - \frac{\left|N_i^{(1)}(t) \cap N_i^{(2)}(t)\right|}{\left|N_i^{(1)}(t) \cup N_i^{(2)}(t)\right|}. \tag{1}$$

The outcome satisfies $0 \le D_i(t) \le 1$ for eligible observations. A value of zero indicates unchanged neighbor relations, whereas one indicates no shared labeled neighbors. **Larger values therefore represent faster or more extensive local structural change, not temporal deceleration.** The code also calculated a turnover rate and its within-anchor relative normalization:

$$\nu_i(t) = \frac{D_i(t)}{\Delta}, \qquad \alpha_i(t) = \frac{\nu_i(t)}{\mathrm{median}_j \nu_j(t)}, \tag{2}$$

with $\alpha_i(t)$ defined only when the anchor median was positive. Additional secondary measures included neighbor novelty and induced ego-edge turnover. The primary fixed-effects models for RQ1, RQ2, and RQ6 used $D_i(t)$, not $\alpha_i(t)$, as their dependent variable.

### 2.4. Time-respecting paths and structural exposure

Within $W_0$, a temporal path from source $j$ to target $i$ was a sequence of events whose timestamps were strictly increasing and whose consecutive endpoints formed a valid walk respecting network directionality. For each reachable ordered pair, the primary temporal duration $d_{ji}^T(t)$ was the smallest difference between the last and first event timestamps among admissible time-respecting paths. An admissible direct one-event path therefore had temporal duration zero. We transformed durations into a dimensionless temporal distance using the network's base window width:

$$\tilde{d}_{ji}(t) = 1 + \frac{d_{ji}^T(t)}{\Delta}. \tag{3}$$

The reference exposure for target $i$ aggregated the mass of all other sources that could reach $i$ temporally:

$$\Phi_i(t) = \sum_{\substack{j \neq i \\ j \rightsquigarrow i}} M_j(t)\, K\left(\tilde{d}_{ji}(t)\right), \qquad K(d) = d^{-2}. \tag{4}$$

Here $M_j(t)$ was the number of distinct neighbors of $j$ in the exposure-window aggregated network, or the sum of its distinct in-neighbor and out-neighbor counts for directed networks. Self-exposure was excluded, and unreachable sources contributed zero. The distance argument in this reference inverse-square kernel was a **normalized temporal-path duration** and should not be interpreted as a geographic, Euclidean, or static shortest-path distance. As the primary path starts and ends within a single window, finite values lie within a narrow domain: $1 \le \tilde{d}_{ji}(t) < 2$. Two additional time-respecting distance constructions, earliest-arrival delay from the start of the exposure window and minimum temporal hop count, were implemented as RQ1 sensitivity analyses.

### 2.5. RQ1: Primary association with subsequent neighborhood stability

The reference exposure was log-transformed and standardized separately within the analyzed network:

$$X_i(t) = \frac{\log(1+\Phi_i(t)) - \overline{\log(1+\Phi)}}{\mathrm{sd}[\log(1+\Phi)]}. \tag{5}$$

The primary model related later turnover to this standardized exposure, common exposure-window controls, node fixed effects, and temporal-anchor fixed effects:

$$D_i(t) = \beta_1 X_i(t) + \boldsymbol{\gamma}^{\mathsf{T}} \mathbf{C}_i(t) + \mu_i + \tau_t + \varepsilon_i(t). \tag{6}$$

The control vector $\mathbf{C}_i(t)$ comprised exposure-window degree, event activity, burstiness, an indicator for insufficient positive inter-event intervals to estimate burstiness, and local clustering. Event activity counted incident events. Burstiness used positive inter-event intervals from the preceding four-window history and was calculated from their sample standard deviation and mean:

$$B_i(t) = \frac{\sigma_i - \mu_i^{\mathrm{IEI}}}{\sigma_i + \mu_i^{\mathrm{IEI}}}, \tag{7}$$

where an unavailable value was replaced by zero and flagged separately when fewer than five strictly positive intervals were available. Fixed effects were absorbed by alternating group demeaning before ordinary least-squares estimation. Under the original directional *temporal deceleration* conjecture, higher exposure would have been associated with lower subsequent turnover ($\beta_1 < 0$). We retained the observed sign of each estimate rather than treating any statistically significant association as support for that conjecture. A fractional-logit sensitivity model and models using secondary turnover outcomes were also implemented; the fractional-logit model did not reproduce the high-dimensional fixed effects and was therefore not used as the primary estimator.

### 2.6. RQ2: High-mass proximity as a stability indicator

For each exposure-window aggregate, high-mass source nodes were those at or above the 90th percentile of degree-based mass. For every eligible target, the nearest reachable high-mass source was selected using $\tilde{d}$; ties were resolved deterministically using source mass and node identifier. The distance of the selected source was then expressed as a within-anchor fractional rank rather than an absolute temporal duration. If $r_i(t)$ denotes its average rank among the $n_t$ targets with observed distances at anchor $t$, the predictor was

$$\rho_i(t) = \frac{r_i(t) - 1}{n_t - 1}, \qquad n_t > 1. \tag{8}$$

The primary model included this relative distance, the standardized log mass of the selected source, the common controls, and node, anchor, and selected-source fixed effects:

$$D_i(t) = \beta_2 \rho_i(t) + \eta X_i^{\mathrm{source}}(t) + \boldsymbol{\gamma}^{\mathsf{T}} \mathbf{C}_i(t) + \mu_i + \tau_t + \psi_{s(i,t)} + \varepsilon_i(t). \tag{9}$$

The value of $\rho_i(t)$ is between zero and one and is **not** measured in seconds, hops, or meters. Supplementary configurations used high-mass cutoffs at the 80th and 95th percentiles, middle-mass pseudo-sources, monotonic adjusted profiles, and pairwise source–target analyses. A coefficient was unavailable if there were insufficient usable observations or no valid high-mass temporal connection.

### 2.7. RQ3: Stability of the exposure–response shape

RQ3 compared six predefined decreasing response-shape families: linear (LIN), exponential (EXP), reciprocal (REC), Hill-type (HILL), square-root (SQ), and monotone nonparametric (MON). For this comparison, log-transformed exposure was mapped onto a network-specific unit interval using its fifth and ninety-fifth percentiles; observations outside that empirical percentile range were excluded from primary curve fitting:

$$x_i(t) = \frac{L_i(t) - Q_{0.05}(L)}{Q_{0.95}(L) - Q_{0.05}(L)}, \qquad L_i(t) = \log\big(1 + \Phi_i(t)\big). \tag{10}$$

After partialling out the common controls and node/anchor fixed effects, the adjusted turnover curve was binned over this interval. Each temporal block's curve was normalized by its lowest-exposure bin, yielding $A(x)$ for comparison with candidate families. The parametric families included, for example,

$$A_{\mathrm{LIN}}(x) = 1 - \zeta x, \quad A_{\mathrm{EXP}}(x) = e^{-\zeta x}, \quad A_{\mathrm{REC}}(x) = \frac{1}{1+\zeta x}, \tag{11}$$

$$A_{\mathrm{HILL}}(x) = c + \frac{1-c}{1+(x/\theta)^q}, \qquad A_{\mathrm{SQ}}(x) = \sqrt{\max(1-\kappa x,\, 10^{-12})}. \tag{12}$$

The nonparametric MON comparator was obtained by decreasing isotonic regression. Parameters were fitted on the training blocks and evaluated against independently constructed adjusted curves for held-out contiguous temporal blocks. Fold losses were weighted by the number of observations in each curve bin. Thus, the RQ3 criterion was **out-of-block MSE of normalized adjusted curves**, not the raw node-level predictive MSE used in RQ4 and RQ5. Comparing their numerical error magnitudes directly would be inappropriate. A one-standard-error rule with an explicit complexity ordering supported across-network candidate selection.

### 2.8. RQ4: Sensitivity to temporal-distance kernels

RQ4 kept the reference degree mass fixed while varying the distance kernel in the exposure sum. The candidate families were a flat kernel (FLAT), inverse-square (ISQ), free power-law (POW), exponential (EXP), reciprocal (REC), stretched exponential (SE), and a constrained nonincreasing binned kernel (MON). The parametric candidates were

$$K_{\mathrm{FLAT}}(d) = 1, \quad K_{\mathrm{ISQ}}(d) = d^{-2}, \quad K_{\mathrm{POW}}(d) = d^{-p}, \tag{13}$$

$$K_{\mathrm{EXP}}(d) = e^{-\xi(d-1)}, \quad K_{\mathrm{REC}}(d) = \frac{1}{1+\xi(d-1)}, \quad K_{\mathrm{SE}}(d) = e^{-\xi \max(d-1,0)^{\chi}}. \tag{14}$$

Parameterized kernels were evaluated on the grids specified in the study configuration. For each candidate, the exposure was transformed with $\log(1+\Phi)$ and used alongside the common controls and node categorical indicators in a Ridge predictor of *raw* turnover $D$. Up to five contiguous temporal folds were formed; each fold was held out while fitting the model on the remaining folds, with at least three usable folds required. The implementation used negligible positive Ridge regularization ($\lambda = 10^{-8}$), standardized numeric inputs, and one-hot-encoded node indicators. With $\mathcal{V}_f$ the observations in test fold $f$, the evaluation criterion was

$$\mathrm{MSE}_f = \frac{1}{|\mathcal{V}_f|} \sum_{(i,t)\in\mathcal{V}_f} \left[D_i(t) - \widehat{D}_i^{(-f)}(t)\right]^2. \tag{15}$$

Development-network mean MSE values were averaged with equal weight across networks to avoid dominance by the largest datasets. Candidate selection applied a one-standard-error threshold and favored simpler transportable models within that threshold. The flexible binned kernel was a predictive benchmark rather than direct validation of a mechanistic attenuation law. High-mass-source-only variants were retained as sensitivity comparisons and did not determine kernel selection.

### 2.9. RQ5: Sensitivity to structural-mass definitions

RQ5 retained the inverse-square kernel while replacing the source mass $M_j$ with alternative attributes of the exposure-window aggregate. Universal candidate definitions were distinct degree (DEG), event activity (ACT), $k$-core index (KS), PageRank (PR), betweenness centrality (BC), harmonic closeness (CC), and eigenvector centrality (EC). Equal mass (EQ) served as a noninformative reference. HITS hub and authority scores were assessed as secondary candidates in directed networks, and native edge-weight strength was examined where weights were meaningful. The same temporally blocked Ridge framework and raw-turnover MSE as RQ4 were used for comparisons.

A nonnegative composite mass (COMP) was also evaluated. To prevent the target network from determining its own composite weights, candidate weights for each target development network were trained on the remaining development networks and then tested on that target. Final weights, when needed for subsequent analyses, were obtained only after candidate selection. RQ5 selection was based on predictive fit and the configured parsimony rule; it did not imply that any network attribute constituted a physically validated mass.

### 2.10. RQ6: Changes in exposure and later neighborhood turnover

RQ6 required an additional preceding exposure window $W_{-1}(t)$, extending from $t - 2\Delta$ (inclusive) to $t - \Delta$ (exclusive), followed by the current exposure window $W_0$ and the two outcome windows $W_1$ and $W_2$. Eligible nodes were active across all four windows. The mass-induced component of exposure change was calculated by substituting the *difference in source mass* into the current exposure geometry:

$$\Delta\Phi_i^{\text{mass}}(t) = \sum_{\substack{j \neq i \\ j \rightsquigarrow i}} \left[M_j^{(0)}(t) - M_j^{(-1)}(t)\right] K\left(\tilde{d}_{ji}^{(0)}(t)\right). \tag{16}$$

Thus, reachability and temporal distances were held at their current-window values when computing the mass-only change; $\Delta\Phi_i^{\text{mass}}(t)$ is **not** the difference between two complete exposure fields computed on independently changing temporal graphs. The fitted association model was

$$D_i(t) = \beta_6 Z_i^{\Delta\Phi}(t) + \lambda Z_i^{\Phi,\text{pre}}(t) + \boldsymbol{\delta}^\top \mathbf{C}_i^{\text{pre}}(t) + \mu_i + \tau_t + \varepsilon_i(t), \tag{17}$$

where $Z_i^{\Delta\Phi}(t)$ was the within-network standardized mass-induced change, $Z_i^{\Phi,\text{pre}}(t)$ was the standardized $\log(1 + \Phi)$ from the previous window, and $\mathbf{C}_i^{\text{pre}}(t)$ consisted of previous-window degree, activity, burstiness, its missingness indicator, and clustering. Supplementary fits separately represented positive and negative magnitudes of mass change and used outcome lags $-4$, $-3$, $0$, $+1$, and $+2$, applying eligibility checks separately to each lag. Because the temporal ordering and measured covariates cannot remove all shared causes, these models were interpreted as associations rather than causal evidence.

### 2.11. Inference, between-network heterogeneity, and multiplicity

For fixed-effects regressions, the implementation first attempted covariance estimation clustered on both node and temporal anchor. If the robust covariance calculation failed, it reverted to the ordinary model covariance *without preserving a fallback indicator in the archived coefficient tables*. Therefore, the available coefficient archive does not establish that two-way clustered standard errors were successfully used for every network. Reported normal-theory confidence intervals and p-values were interpreted subject to this unresolved implementation-level uncertainty; this limitation was not concealed by treating the nominal inference as uniformly robust.

For estimable network-level coefficients with finite positive standard errors, the configured synthesis used the DerSimonian–Laird random-effects method (DerSimonian and Laird, 1986). If $\hat{\beta}_k$ and $s_k$ represent the estimate and standard error from network $k$, the random-effects summary was

$$\hat{\beta}_{\text{RE}} = \frac{\sum_{k=1}^{K} w_k \hat{\beta}_k}{\sum_{k=1}^{K} w_k}, \qquad w_k = \frac{1}{s_k^2 + \hat{\tau}_{\text{DL}}^2}. \tag{18}$$

The method estimated between-network variance $\hat{\tau}_{\text{DL}}^2$ and the Cochran $Q$ statistic; heterogeneity was reported alongside the pooled coefficient. Each network contributed one coefficient to the synthesis rather than contributing proportionally to its observation count. Nominal two-sided normal-theory intervals were used. As an explicitly *post hoc* secondary analysis of the archived coefficient estimates, Benjamini–Hochberg false-discovery-rate adjustment (Benjamini and Hochberg, 1995) was applied within each research-question family. Undefined estimates, zero standard errors, and nonestimable models were excluded rather than counted as evidence of a zero effect. In view of strong across-network variation and the covariance-fallback limitation, the network-specific effect sizes and their directions were central to interpretation.

### 2.12. Temporal null models and treatment of incomplete computations

Four core temporal randomization families were implemented: global timestamp shuffling, edgewise inter-event-interval reordering, activity-preserving partner randomization, and degree-preserving temporal rewiring. Each randomized a different aspect of event timing, partner identity, or local interaction structure while preserving selected properties of the source stream. The RQ1 null workflow reused the same turnover definition, exposure transformation, covariates, and fixed-effects estimation routine as the corresponding empirical model. Randomization could nevertheless change the nodes satisfying the three-window eligibility rule and therefore change the analyzed node–anchor samples. The null results were interpreted as *procedure-specific reference distributions* rather than identically matched-sample causal counterfactuals.

The configured null design comprised 200 realizations for each of four families in each of 12 development networks (9,600 planned realizations). The available archived output contained 3,606 core realization-level JSON records; other planned realizations were not treated as completed. Large-scale computational runs, including a portion of the NYC Yellow Taxi null simulations, had been stopped because of computational cost. Completed primary analyses for such networks were retained, while missing null realizations were recorded as unobserved rather than negative findings. The repository also contained code for synthetic validation and the reserved replication panel, but the reviewed result archive contained no numerical outputs that would justify claiming completed synthetic validation or independent replication.

### 2.13. Computational reproducibility and interpretation scope

The examined materials consisted of the available output tables, source modules, dataset and study configurations, protocol-amendment records, and software tests. Raw source event streams, complete normalized caches, all run-state artifacts, and a single frozen manifest tying every estimate to an immutable software revision were not available in the reviewed archive. We therefore report a source-to-output methodological consistency assessment, **not** an independent rerun of every recorded result from raw data. Version labels differed among repository files, and one legacy test referred to functions no longer present in the archived null-runner interface; these packaging inconsistencies do not alter the reported coefficients but limit a claim of end-to-end reproducibility in the supplied snapshot. Taken together, the design supports a transparent comparative analysis of completed observational results across the 12 development networks. It does not establish a universal physical time law, a causal process, or independently replicated generalization outside that panel.

## 3. Results

### 3.1. Availability of the empirical analyses

The completed development-panel results covered 12 heterogeneous temporal networks (Table 1). Primary exposure–turnover coefficients (RQ1) were estimable in 11 networks, proximity-to-high-mass coefficients (RQ2) in 10, and mass-change coefficients (RQ6) in eight. Bitcoin OTC yielded an estimated coefficient of zero with zero standard error and an undefined p-value in all three corresponding outputs, and Facebook Wall yielded the same nonestimable pattern for RQ2; these records were not interpreted as evidence of zero association. Other RQ6 network outputs were unavailable rather than statistically nonsignificant. All 12 development networks contributed cross-validation summaries for RQ3–RQ5, subject to the qualification that Bitcoin OTC had zero reported curve and predictive losses, limiting the information carried by this case.

*Table 2. Availability of primary empirical results and completed core temporal-randomization records.*

| Development network | RQ1 | RQ2 | RQ6 | RQ3–RQ5 CV | Core nulls |
|---|---|---|---|---|---|
| High School 2013 | ✓ | ✓ | ✓ | ✓ | 400/800 |
| Workplace 2015 | ✓ | ✓ | ✓ | ✓ | 403/800 |
| SFHH Conference | ✓ | ✓ | ✓ | ✓ | 399/800 |
| Email-EU-core | ✓ | ✓ | — | ✓ | 398/800 |
| CollegeMsg | ✓ | ✓ | ✓ | ✓ | 405/800 |
| Manufacturing E-mail | ✓ | ✓ | ✓ | ✓ | 399/800 |
| Facebook Wall | ✓ | NE | — | ✓ | 404/800 |
| MathOverflow | ✓ | ✓ | — | ✓ | 399/800 |
| Bitcoin OTC | NE | NE | NE | ✓ | 396/800 |
| NYC Yellow Taxi | ✓ | ✓ | ✓ | ✓ | 3/800 |
| NYC Citi Bike | ✓ | ✓ | ✓ | ✓ | 0/800 |
| Chicago Divvy | ✓ | ✓ | ✓ | ✓ | 0/800 |

Note. ✓, an archived estimable primary coefficient (RQ1, RQ2, or RQ6) or completed comparative cross-validation summary (RQ3–RQ5); NE, zero standard error and undefined primary p-value; —, no primary RQ6

output. Null-model counts indicate saved realization files, not tests passing or failing. The design scheduled 800 core realizations per network (four null families × 200).

Across the four null families and 12 development networks, 3,606 of 9,600 planned core randomizations (37.6%) had realization-level JSON outputs. The distribution of completed runs was uneven: NYC Yellow Taxi had three, whereas NYC Citi Bike and Chicago Divvy had none. These computational shortfalls were handled as unavailable results, not negative findings. No numerical outputs from the reserved independent replication panel or planned synthetic validation were present in the reviewed archive.

### 3.2. RQ1: Directional consistency of structural exposure as a stability indicator

Of 11 estimable network-specific associations between standardized log exposure and subsequent Jaccard neighborhood turnover, two positive coefficients survived the post hoc within-RQ Benjamini–Hochberg false-discovery-rate (BH-FDR) correction: High School 2013 ($\beta = +0.03424$, 95% CI [+0.02112, +0.04735], $q < 0.0001$; $n = 19{,}291$) and SFHH Conference ($\beta = +0.04691$, 95% CI [+0.03011, +0.06370], $q < 0.0001$; $n = 7{,}656$). These associations run opposite to the original deceleration conjecture, which predicted lower turnover at higher exposure ($\beta < 0$).

Email-EU-core ($\beta = -0.01485$, nominal $p = 0.0286$) and NYC Citi Bike ($\beta = -0.00024$, nominal $p = 0.0495$) had negative nominal associations, but neither survived BH-FDR correction ($q = 0.1050$ and $0.1292$, respectively). The remaining seven estimable networks did not meet the corrected significance threshold. Thus, the available data do not establish a consistently negative exposure–turnover association across the development panel. Effect magnitudes also varied substantially, from negligible values in several mobility networks to larger positive coefficients in two face-to-face contact networks (Figure 1; Table S1).

Figure 1. RQ1 network-specific standardized-exposure coefficients and nominal 95% confidence intervals. Filled markers identify BH-FDR $q < 0.05$; hollow markers indicate $q \geq 0.05$. The vertical dashed line denotes zero. Bitcoin OTC was excluded because the primary estimate was nonestimable. Negative coefficients correspond to higher subsequent stability (lower turnover); positive coefficients correspond to lower stability. The historical prespecified conjecture anticipated a negative coefficient.

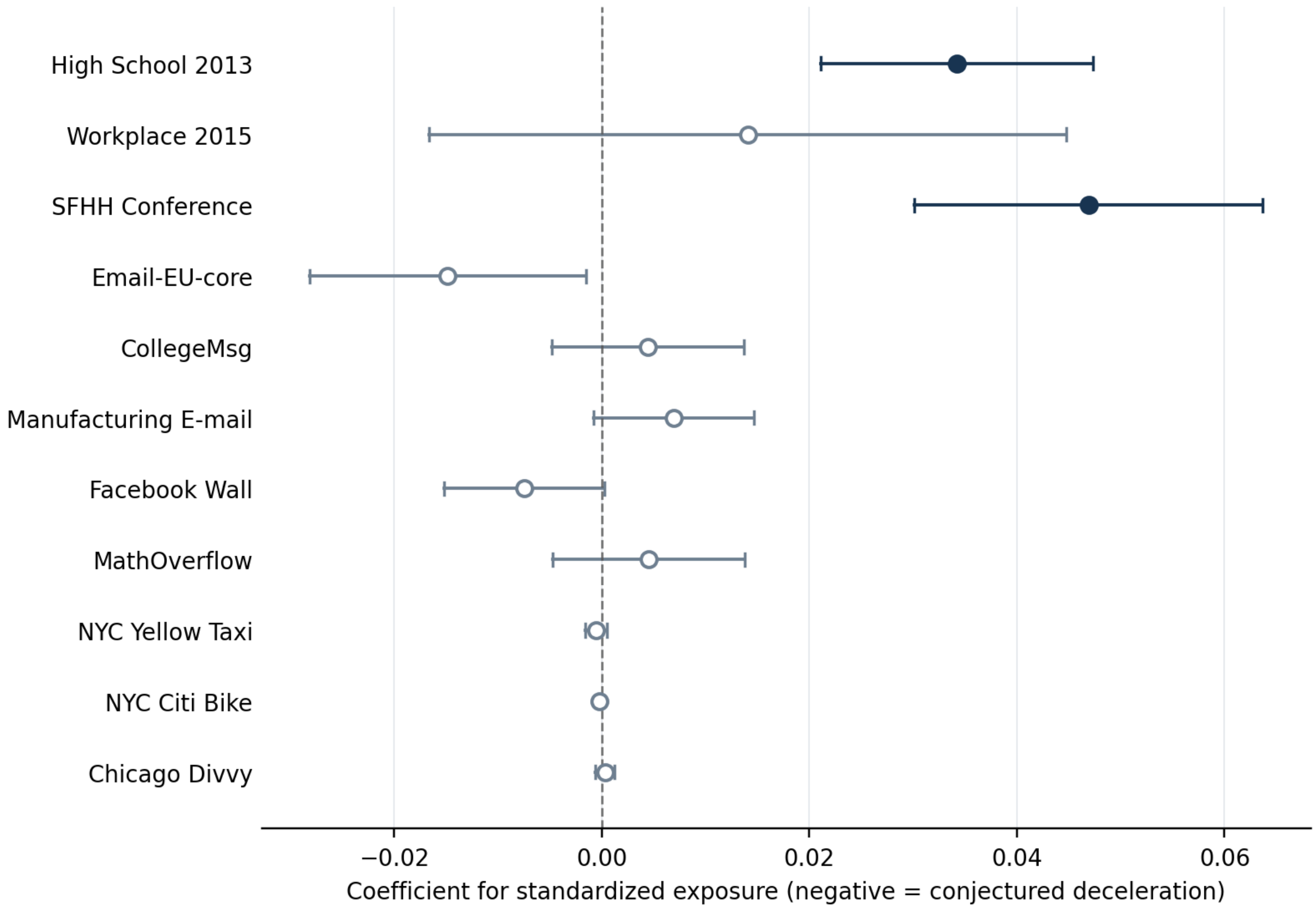

### 3.3. RQ2: Stability gradients near high-mass sources

The high-mass proximity coefficient was estimable in 10 development networks. NYC Yellow Taxi was the only network retaining significance after within-RQ BH-FDR correction ($\beta = +0.00873$, 95% CI [+0.00430, +0.01316], nominal $p = 0.000114$, $q = 0.00114$; $n = 586{,}496$). NYC Citi Bike showed a smaller nominal positive coefficient ($\beta$ = +0.000924, nominal p = 0.0263) that did not survive adjustment (q = 0.1317). None of the other eight estimable networks met $q < 0.05$ (Table S1). Because the predictor is a within-anchor fractional rank of distance to the closest temporally reachable high-mass source, these estimates concern relative rank, not elapsed seconds or hop counts.

The archived DerSimonian–Laird random-effects synthesis across the 10 estimable networks yielded $\beta$ = +0.002637 (95% CI [-0.001189, +0.006463]). The interval included zero, with Q = 15.27 (9 degrees of freedom) and $I^2 \approx 41.0\%$. Consequently, the panel did not provide clear evidence of a common, transportable proximity gradient, although the NYC Yellow Taxi association was detectably positive under the archived inference procedure.

### 3.4. RQ3: Transferability of the adjusted exposure–response shape

The six prespecified decreasing function families were compared by out-of-block mean squared error (MSE) on normalized, control-adjusted turnover curves (72 network–family summaries). The smallest equal-network mean MSE was obtained by the Hill family (0.002181). Other parametric families differed only slightly at the aggregate level, whereas the monotone nonparametric (MON) family had the highest mean MSE (0.002274) even though it achieved the lowest network-level loss in six of the 12 networks. Square-root (SQ) had the smallest loss in three networks; HILL, reciprocal (REC), and exponential (EXP) each in one; linear (LIN) in none. Bitcoin OTC contributed a reported zero loss for all families; its tie was resolved by the archive row ordering in the winner tally and should not be interpreted as substantive evidence favoring MON.

*Table 3. Cross-validated RQ3 response-shape comparisons across 12 development networks.*

| Family | Response model | Mean CV MSE | Network SE | Lowest-MSE networks |
|---|---|---|---|---|
| HILL | Hill-type | 0.002181 | 0.000812 | 1 |
| SQ | Square root | 0.002214 | 0.000826 | 3 |
| LIN | Linear | 0.002216 | 0.000827 | 0 |
| EXP | Exponential | 0.002218 | 0.000827 | 1 |
| REC | Reciprocal | 0.002219 | 0.000828 | 1 |
| MON | Monotone nonparametric | 0.002274 | 0.000853 | 6 |

Note. CV MSE measures prediction of normalized, adjusted binned curves, not raw node-level turnover. The network counts use the recorded minimum per network; the Bitcoin OTC all-zero tie is noninformative. Standard errors describe across-network dispersion, not formal pairwise model-comparison tests.

The distinction between smallest mean error and most frequent within-network minimum prevents identification of a unique response law. Moreover, all candidate response families were constrained to be nonincreasing; a low curve-fitting error alone does not independently confirm the originally hypothesized negative association, especially in networks where the primary RQ1 coefficient was positive.

### 3.5. RQ4: Predictive sensitivity to distance kernels

Across 55 candidate distance kernels evaluated using blocked cross-validation of raw turnover, the lowest equal-network mean MSE was observed for SE_xi4.0_chi0.5 (0.058223), compared with 0.058304 for FLAT and 0.058278 for the reference inverse-square kernel (ISQ). The best grid candidate reduced aggregate mean MSE relative to FLAT by only 0.14%. All 55 candidate kernels were marked as falling within the archive's one-standard-error admissibility region. This is a model-selection rule, not an equivalence test; the results therefore do not resolve a uniquely preferred decay law.

### 3.6. RQ5: Predictive sensitivity to structural-mass representations

The PageRank mass candidate (PR) achieved the smallest equal-network mean raw-turnover CV MSE among eight recorded structural-mass candidates (0.058142), compared with 0.058278 for distinct degree (DEG); the relative difference was 0.23% of the DEG mean. PR had the smallest recorded loss in seven networks; in other networks, KS, BC, CC, or EC was lowest, while Bitcoin OTC showed a noninformative all-zero tie among the candidates. All eight archived mass candidates were within the one-standard-error region, so this selected model cannot be presented as a uniquely identified physical mass.

*Table 4. RQ4 and RQ5 mean prediction errors, selected comparators, and admissibility of competing models.*

| Comparison | Candidate | Mean CV MSE | Selection caveat |
|---|---|---|---|
| RQ4: best grid kernel | SE_xi4.0_chi0.5 | 0.058223 | 55/55 within 1 SE |
| RQ4: flat reference | FLAT | 0.058304 | Reference |
| RQ4: inverse-square | ISQ | 0.058278 | Reference kernel |
| RQ5: smallest mean MSE | PR | 0.058142 | 8/8 within 1 SE |
| RQ5: degree reference | DEG | 0.058278 | Reference mass |

Note. RQ4 and RQ5 both predict raw Jaccard turnover using temporally blocked Ridge models, so their losses are on the same outcome scale; the RQ3 curve-normalization loss is not numerically comparable. "Within 1 SE" is the archive's across-network candidate-selection flag and does not demonstrate equality of model performance. No independent held-out replication result was archived for the selected PR and ISQ components.

### 3.7. RQ6: Directional consistency of mass-induced exposure changes

Eight networks yielded estimable associations between standardized mass-induced changes in structural exposure and later neighborhood turnover. Four associations remained significant after within-RQ BH-FDR correction. High School 2013 ($\beta$ = +0.02526, 95% CI [+0.01238, +0.03814], q = 0.00049) and Workplace 2015 ($\beta$ = +0.05199, 95% CI [+0.00760, +0.09638], q = 0.0434) had positive associations. NYC Yellow Taxi ($\beta$ = −0.02513, 95% CI [−0.02958, −0.02068], q < 0.0001) and NYC Citi Bike ($\beta$ = −0.000379, 95% CI [−0.000617, −0.000141], q = 0.00474) had negative associations. The contrast in sign was not explained away by treating nonsignificant individual coefficients as zero (Figure 2; Table S1).

The archived random-effects synthesis gave $\beta$ = -0.002841 (95% CI [-0.005746, +0.000064]); the interval crossed zero. Between-network heterogeneity was extreme (Q = 143.46, 7 degrees of freedom; $I^2 \approx 95.1\%$). The pooled coefficient should thus be read alongside, not instead of, the divergent network estimates. Because RQ6 fixes current-window reachability and path geometry when computing the mass-induced component of exposure change, these estimates do not establish the effect of a full change in temporal-network geometry or a causal temporal mechanism.

*Figure 2. RQ6 network-specific coefficients, nominal 95% confidence intervals, and archived random-effects summary. Filled points indicate BH-FDR q < 0.05; the diamond represents the pooled coefficient. The pooled interval crosses zero despite strong and opposite-signed network-level associations.*

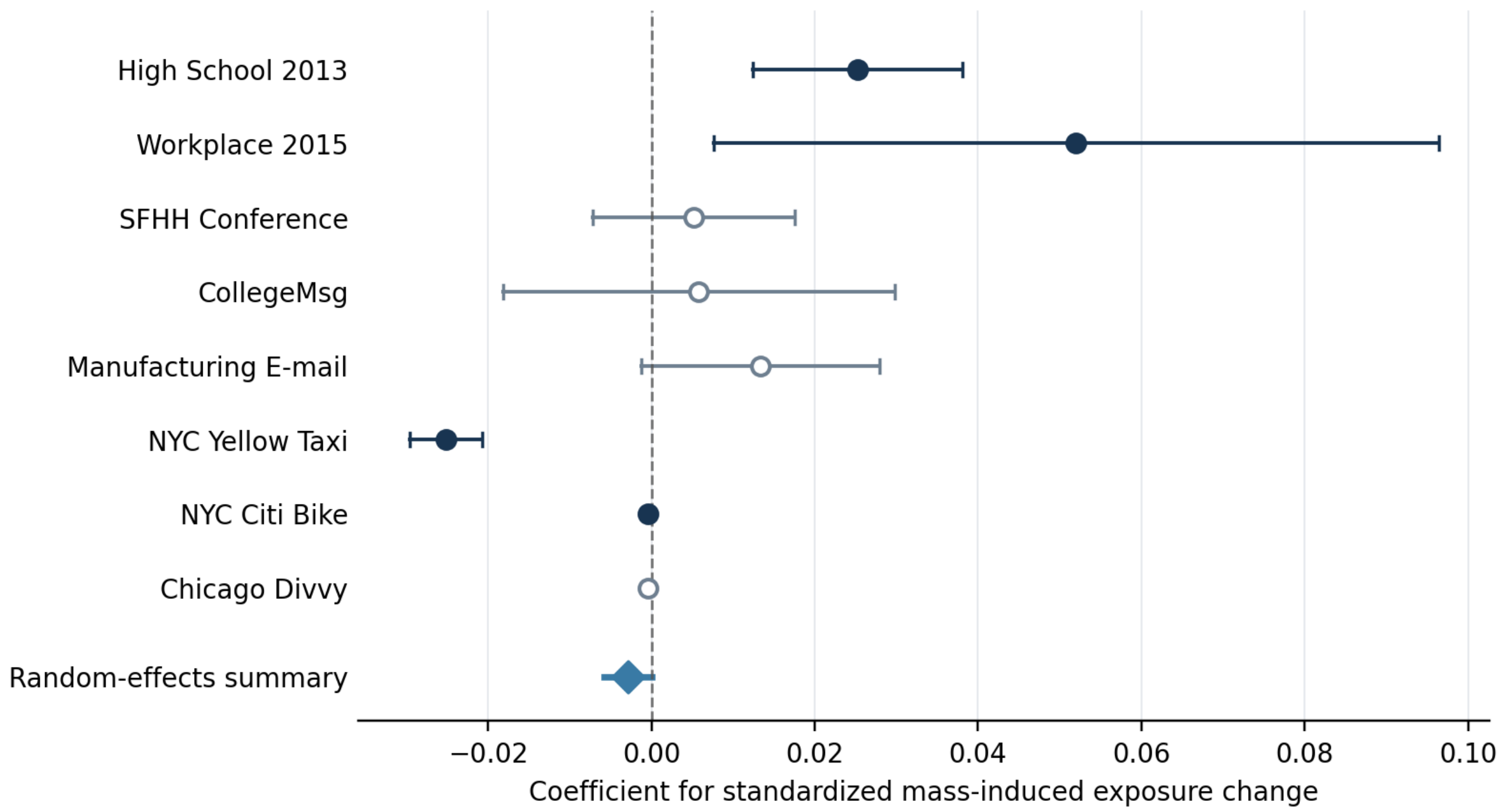

### 3.8. Temporal randomization: incomplete but informative sensitivity evidence

Four temporal randomization families generated 3,606 saved core realization records, with heterogeneous coverage among network–family combinations. The RQ1 randomization pipeline reused the empirical turnover definition, exposure transformation, controls, and fixed-effects estimator. However, randomized event streams changed the number of node–anchor observations meeting the three-window eligibility rule; the null coefficients therefore form procedure-specific reference distributions and are not derived from identical matched observation panels. Given the incomplete realization matrix and the heterogeneous null-preservation properties, comparisons are reported as exploratory distributional diagnostics rather than confirmatory p-values.

*Table 5. RQ1 descriptive comparisons with available null-model realizations in the two networks with positive BH-FDR-significant primary effects.*

| Network | Null family | n | Observed β | Null mean β | Upper-tail share |
|---|---|---|---|---|---|
| High School | Partner randomization | 99 | +0.03424 | -0.00106 | 0.010 |
| High School | Temporal rewiring | 98 | +0.03424 | -0.00092 | 0.010 |
| High School | Edgewise shuffle | 102 | +0.03424 | +0.02787 | 0.097 |
| High School | Global shuffle | 101 | +0.03424 | +0.00193 | 0.010 |
| SFHH | Partner randomization | 101 | +0.04691 | -0.00179 | 0.010 |
| SFHH | Temporal rewiring | 100 | +0.04691 | -0.00213 | 0.010 |
| SFHH | Edgewise shuffle | 97 | +0.04691 | +0.06568 | 1.000 |
| SFHH | Global shuffle | 101 | +0.04691 | +0.00417 | 0.010 |

Note. Upper-tail share = (number of saved null coefficients at least as large as the observed coefficient + 1)/(number of valid saved null coefficients + 1). It is reported to describe the archived reference distribution only, not as a confirmatory test controlling the multiplicity of networks and null families. The null sample can differ from the empirical sample after eligibility filtering.

For High School 2013, the observed RQ1 coefficient (+0.03424) exceeded the central tendency of activity-preserving partner randomization, degree-preserving rewiring, and global timestamp shuffling, but was much closer to the edgewise-shuffle distribution (mean +0.02787; exploratory upper-tail share 0.097). For SFHH, the observed coefficient (+0.04691) likewise exceeded the central tendency of the first three references but was below the edgewise-shuffle mean (+0.06568); its exploratory lower-tail share was approximately 0.010. Figure 3 displays the different distributions. Thus, the evidence that an empirical coefficient differs from a randomization reference depends strongly on which temporal or structural features the null family preserves.

*Figure 3. Descriptive RQ1 null-coefficient distributions for (a) High School 2013 and (b) SFHH. The dashed line is the empirical coefficient; horizontal bars show the interquartile range and whiskers the 5th–95th percentile of completed null coefficients. Comparisons use unequal post-randomization eligible samples and are not confirmatory hypothesis tests.*

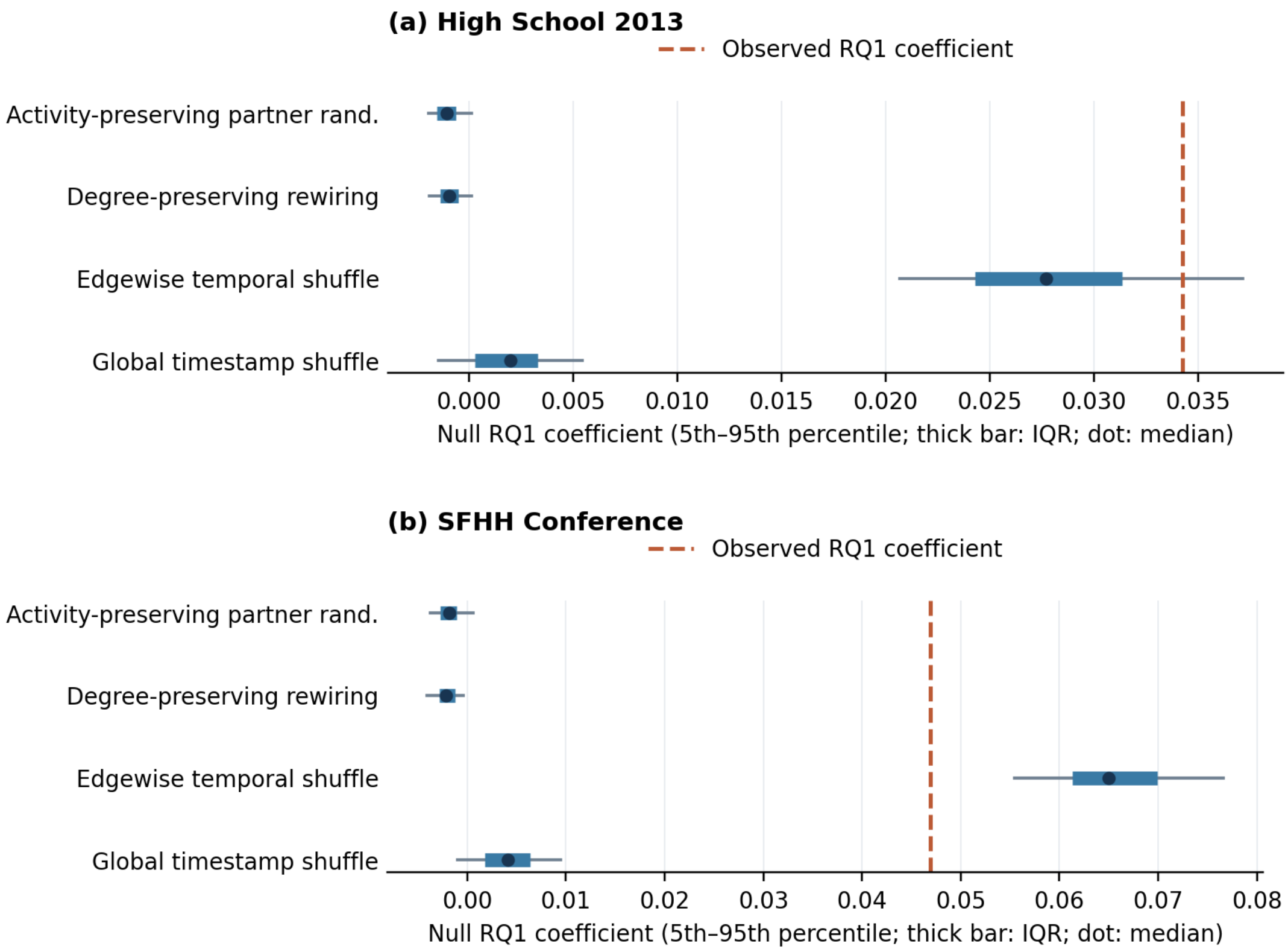

### 3.9. What the completed evidence establishes about indicator reliability

The completed results do not establish a directionally consistent relation between temporal structural exposure and later neighborhood stability across the 12 development networks. The only RQ1 associations remaining significant after the post hoc within-question adjustment were positive on the turnover scale, implying lower subsequent stability rather than the originally expected greater stability. The pooled RQ2 gradient crossed zero, while the RQ6 analysis contained significant effects in opposing directions. The RQ3–RQ5 comparisons show little separation among candidate specifications under their respective cross-validation losses; they do not test whether including exposure improves predictions over an otherwise identical model without exposure. Null analyses were incomplete, and covariance-fallback status was not archived. Together, the findings argue against a network-independent, universally directional exposure–stability indicator under this operational construction, without proving that structural exposure has no predictive or scientific value in individual domains.

## 4. Discussion

### 4.1. What would count as a reliable indicator?

The central question of this study is whether structural exposure constructed from source mass and temporally feasible distances is a reliable indicator of subsequent local neighborhood stability. An answer requires distinguishing statistical association, consistency of effect direction, robustness of the exposure definition, and incremental predictive value. The archived evidence is strongest for a negative conclusion about universal directional consistency: the primary exposure–turnover relation does not retain one sign across the development networks, and the dynamic mass-change associations are also directionally heterogeneous. These observations do not mean that topology is irrelevant, nor that a useful indicator cannot exist for a narrower domain or timescale.

A stronger claim of predictive reliability would require an explicit predictor with exposure to outperform an otherwise comparable predictor without exposure on held-out periods, preferably followed by evaluation on unseen networks. RQ4 and RQ5 select among exposure representations but do not provide that particular no-exposure benchmark; reserved replication-network results were unavailable. Accordingly, we interpret "reliable" in the restricted sense of cross-network association direction and sensitivity to operational choices, not as demonstrated out-of-sample forecasting skill. This is a substantive boundary on what the article establishes, not merely a wording preference.

### 4.2. Directional inconsistency and the meaning of positive turnover coefficients

The primary response is Jaccard neighborhood turnover: more turnover means less local structural stability. Thus, a positive exposure–turnover coefficient should not be labeled "stabilization." High School 2013 ($\beta = +0.03424$) and SFHH ($\beta = +0.04691$) retained significance after post hoc within-RQ BH-FDR adjustment, but both indicate *lower* later stability at higher standardized exposure under the fitted models. The originally specified structural-deceleration conjecture predicted a negative turnover coefficient, so these results point in its opposite direction. They provide a falsifying contrast for that conjecture as operationalized, subject to the covariance-estimation caveat discussed below.

The two nominal negative RQ1 effects in Email-EU-core and NYC Citi Bike did not survive the same multiplicity adjustment. Other estimable networks yielded no adjusted significant association. Lack of significance is not equivalence to zero, and because the coefficient standard errors may be affected by unrecorded covariance fallbacks, significance labels themselves should be regarded as provisional. More robust than any single p-value is the descriptive observation that the estimated direction and magnitude are uneven. A universal, one-sign exposure–stability association cannot be inferred from this panel.

An exposure field and a local stability outcome need not move together. Exposure summarizes access to masses through temporally valid routes; turnover counts replacement of immediate labeled neighbors in future windows. A node may be temporally close to prominent sources while maintaining a fixed partner set, or while encountering a changing collection of partners. Prior work establishes the distinction between static and temporally feasible paths (Pan and Saramäki, 2011), while the temporal gravity model of Bi et al. (2021) addresses node influence, not neighborhood persistence. Our findings therefore delimit an outcome-specific use of mass–distance exposure rather than discrediting that modeling family for other purposes.

### 4.3. Model robustness is not the same as validated predictive usefulness

RQ2 did not reveal a common high-mass proximity gradient: the pooled coefficient was +0.002637, with a 95% confidence interval of [−0.001189, +0.006463], and heterogeneity was non-negligible ($I^2 \approx 41\%$). NYC Yellow Taxi was the only network with an adjusted significant proximity association. Because the predictor was ranked distance to a reachable high-mass source, this result is not interpretable as a universal effect per unit of temporal duration. The absence of a clear pooled gradient weakens the case for a single proximity rule but does not exclude context-specific gradients.

In RQ3, HILL had the smallest equal-network mean loss on normalized adjusted response curves, while MON had the lowest loss in six networks. These observations do not identify one preferred exposure–stability law. Crucially, all RQ3 candidates were constrained to decreasing responses; their fit cannot demonstrate that the empirical relation is decreasing when RQ1 finds positive coefficients in prominent cases. The RQ3 loss is also defined on binned, adjusted curves and cannot be directly compared numerically with the raw-turnover predictive losses of RQ4 and RQ5.

In RQ4, the best candidate kernel reduced equal-network mean raw-turnover MSE by about 0.14% relative to the flat reference. In RQ5, PageRank mass improved the corresponding mean by about 0.23% relative to degree mass. All archived candidates in both comparisons fell within their respective one-standard-error selection regions. These statements support modest differentiation *among* exposure specifications, not statistical equivalence, and certainly not independent demonstration of incremental predictive usefulness over a model that omits exposure altogether.

Accordingly, the study supports a limited decision rule for method developers: do not adopt a complex mass–distance exposure indicator merely because it outperforms another exposure variant by a small average MSE. Report absolute errors, simpler exposure specifications, a true no-exposure baseline if it has been evaluated, and uncertainty in model selection. The present archive lacks that full comparative chain and should not be described as completing it.

### 4.4. Exposure changes and instability of association direction

RQ6 is a separate dynamic analysis of mass-induced exposure change. Four of eight estimable networks retained adjusted significance, split between positive effects for High School 2013 and Workplace 2015 and negative effects for NYC Yellow Taxi and NYC Citi Bike. The pooled coefficient (−0.002841; 95% CI [−0.005746, +0.000064]) crossed zero, whereas $I^2$ was approximately 95.1%. This strong descriptive variation means the pooled average alone is not a defensible summary of indicator direction. It is evidence about an operational mass-change component, not about all conceivable structural change.

The RQ6 predictor substitutes the change in source mass while holding current-window temporal reachability and path geometry fixed. This construction deliberately isolates one component of the exposure field. It does not capture concurrent path rewiring and cannot establish that changes in exposure cause subsequent stability changes. The time ordering improves temporal separation between recorded predictor and response but does not remove unobserved common causes.

### 4.5. Temporal reference models and the definition of the comparison

Reference models can identify which recorded temporal or structural regularities are relevant to an observed association only relative to their own preservation rules. Gauvin et al. (2022) provide a taxonomy of temporal-network randomization procedures, and Hobson et al. (2021) explain general pitfalls in using network reference distributions. In our analysis, global timestamp shuffling, edgewise interval reordering, partner randomization, and degree-preserving temporal rewiring address different questions; they are not alternative implementations of one universal null hypothesis.

For High School 2013, the observed RQ1 coefficient was only moderately above the edgewise-shuffle center; its exploratory upper-tail share was 0.097. In SFHH, the observed coefficient fell below the edgewise-shuffle distribution's mean, even though it exceeded the central tendency of other null families. This contrasts with a simple claim that the observed coefficient is uniquely generated by the original event ordering. It instead shows that the interpretation depends on the features retained by a specific transformation; the procedure cannot identify a unique physical or causal mechanism.

Only 3,606 of the 9,600 scheduled core null realizations were archived. Some large computations were stopped, and mobility coverage was especially sparse. Further, randomization changed the node–anchor observations satisfying the eligibility filters; the empirical and randomized coefficients can be based on different panels. Hence these comparisons are exploratory diagnostics, not confirmatory matched-sample p-values. Incomplete coverage is missing evidence, not a negative result.

### 4.6. Why might the indicator behave differently across networks?

The observed heterogeneity may partly reflect differences in interaction mechanisms, though the present data do not identify specific causes. In bounded face-to-face systems, recurring schedules and copresence constrain possible contacts. Emails and online contacts allow asynchronous and geographically unconstrained exchange. Mobility records often reflect repeated travel demand rather than enduring social relationships. An identical scalar exposure construction therefore need not have an identical association with neighbor retention in these settings. Ceria et al. (2022)also found differences in topological–temporal contact correlations across physical and virtual networks, but their contact-level outcome is not equivalent to our node-neighborhood stability outcome.

Resolution and measurement also matter. Neighborhood turnover is computed between two future windows, while exposure is derived from masses and temporal paths in an earlier window; activity thresholds determine which observations appear in the panel. These choices influence both the denominator of the Jaccard comparison and the set of temporally reachable sources. They might help explain observed contrasts, but the study was not designed to isolate the causal impact of each choice. The existence of multiple plausible measurement pathways strengthens the case for stating the scope of the indicator precisely.

Additional explanations may involve temporal memory, higher-order pathways, and differences in event generation. Scholtes (2017) and Lambiotte et al. (2019) show why pairwise or first-order network representations can miss path-dependent behavior. Zou et al. (2023) examine network memory in temporal prediction, a related but distinct forecasting task. These references justify alternative hypotheses; they do not establish why the coefficients in the present dataset differ in sign.

### 4.7. Statistical and computational boundaries of inference

The most consequential inferential issue is the covariance-estimation fallback: the fixed-effects routine attempted two-way cluster-robust standard errors, but could revert to ordinary covariance without recording when this happened. The reported confidence intervals, BH-FDR labels, and between-network syntheses therefore inherit unresolved uncertainty. We retain the archived values rather than retroactively claiming uniform robustness. The DerSimonian–Laird aggregation summarizes a heterogeneous group of observational estimates, not independent out-of-network transportability, and the post hoc multiplicity analysis was not part of an externally preregistered confirmatory design.

The three modeling comparisons answer distinct questions. RQ3 evaluates normalized adjusted curves among decreasing candidates; RQ4 and RQ5 evaluate raw-turnover Ridge prediction with blocked folds. Neither RQ4 nor RQ5 includes the specific baseline needed to claim that exposure adds predictive value beyond the same controls and node effects. The study also does not present calibration, prospective test performance, or held-out replication networks. Therefore, a strong statement that exposure is *predictively useless* would exceed the evidence just as much as a claim that it is *universally reliable*.

Computational availability was uneven. The primary analyses remain usable where their estimates were archived, but some NYC Yellow Taxi randomizations were terminated because of processing costs; independent replication and synthetic-recovery outputs were unavailable. The examined repository was also not tied to one frozen and fully reproducible software revision. These are substantive restrictions on transportability and reproducibility, not evidence against the proposed indicator under untested conditions.

### 4.8. Implications for research using structural exposure

For researchers designing temporal structural features, the main practical finding is conditional rather than categorical. A temporally reachable, mass-weighted exposure quantity may correlate with later neighborhood turnover in some networks, but its direction and useful specification should not be presumed before domain-specific validation. Reporting the sign and uncertainty of exposure effects, performance relative to a no-exposure baseline, and tests across distinct datasets would make future indicator claims more interpretable. The present study supplies a common observational comparison and an explicit account of where this particular construction lacks consistent directional support.

Our contribution is not a new gravity formulation, a validated time-dilation law, or a cross-network forecasting benchmark. It is a critical, outcome-specific assessment of an intuitively appealing structural indicator under temporally ordered observation. By treating significant positive effects, nonsignificant effects, model-selection ties, and unfinished simulations differently, the analysis provides a more useful empirical boundary than a simple "hypothesis failed" narrative. The implication for future work is to identify settings in which exposure is informative and to demonstrate that information with appropriately matched predictive and validation comparisons.

## 5. Conclusion

This study asked whether temporally weighted structural exposure is a reliable indicator of subsequent local neighborhood stability across heterogeneous temporal networks. Under a common exposure and turnover definition, the available 12-network development panel did not reveal a directionally consistent relation. The two primary associations surviving post hoc within-question false-discovery-rate adjustment were positive on the turnover scale, indicating lower subsequent stability at greater exposure, contrary to the originally anticipated direction. Other network-specific relationships were mixed or uncertain, and mass-induced exposure changes were associated with later turnover in opposite directions across networks.

The functional-form, temporal-distance kernel, and structural-mass comparisons did not identify a transferable specification with a clearly meaningful advantage over simpler exposure alternatives. Available randomized reference results showed that interpretations depend on the properties preserved by each null procedure, while changes in sample eligibility limit exact like-for-like comparisons. These findings challenge the use of the examined exposure construction as an unqualified, network-independent indicator.

The evidence does not establish that structural exposure is intrinsically uninformative or that no exposure-based model could predict future stability. The archived predictive comparisons lack an explicit no-exposure baseline and external replication, core null simulations were incomplete, and covariance-estimation fallbacks were not recorded. Accordingly, our conclusions are limited to direction, heterogeneity, and specification sensitivity in the completed observational analyses; they are neither causal claims nor out-of-network forecasts.

The practical implication is that structural exposure should be treated as a candidate domain-dependent feature whose relationship with neighborhood persistence requires direct validation. Future assessments should test its incremental predictive value, robustness to operational definitions, and transfer across independently held-out networks. Within the scope of the present evidence, a universal exposure–stability rule has not been demonstrated.

## Declaration of Generative AI Use

The author acknowledges the use of OpenAI's ChatGPT as a generative artificial intelligence tool during the preparation of this manuscript. ChatGPT was employed to assist with manuscript drafting and restructuring, academic language editing, identification of relevant literature, preliminary source-code review, and the organization and interpretation of existing experimental results. AI-generated suggestions were treated as supporting material rather than independent scientific evidence. The reported empirical findings are based on the study's computational outputs. The author retains full responsibility for the research methodology, accuracy of the reported results, interpretation of findings, verification of references, and final manuscript content.

## References

Benjamini, Y., & Hochberg, Y. (1995). Controlling the false discovery rate: A practical and powerful approach to multiple testing. Journal of the Royal Statistical Society: Series B (Methodological), 57(1), 289–300. https://doi.org/10.1111/j.2517-6161.1995.tb02031.x

Bi, J., Jin, J., Qu, C., Zhan, X., Wang, G., & Yan, G. (2021). Temporal gravity model for important node identification in temporal networks. Chaos, Solitons & Fractals, 147, 110934. https://doi.org/10.1016/j.chaos.2021.110934

Ceria, A., Havlin, S., Hanjalic, A., & Wang, H. (2022). Topological–temporal properties of evolving networks. Journal of Complex Networks, 10(5), cnac041. https://doi.org/10.1093/comnet/cnac041

DerSimonian, R., & Laird, N. (1986). Meta-analysis in clinical trials. Controlled Clinical Trials, 7(3), 177–188. https://doi.org/10.1016/0197-2456(86)90046-2

Gauvin, L., Génois, M., Karsai, M., Kivelä, M., Takaguchi, T., Valdano, E., & Vestergaard, C. L. (2022). Randomized reference models for temporal networks. SIAM Review, 64(4), 763–830. https://doi.org/10.1137/19M1242252

Hobson, E. A., Silk, M. J., Fefferman, N. H., Larremore, D. B., Rombach, P., Shai, S., & Pinter-Wollman, N. (2021). A guide to choosing and implementing reference models for social network analysis. Biological Reviews, 96(6), 2716–2734. https://doi.org/10.1111/brv.12775

Holme, P., & Saramäki, J. (2012). Temporal networks. Physics Reports, 519(3), 97–125. https://doi.org/10.1016/j.physrep.2012.03.001

Karsai, M., Kivelä, M., Pan, R. K., Kaski, K., Kertész, J., Barabási, A.-L., & Saramäki, J. (2011). Small but slow world: How network topology and burstiness slow down spreading. Physical Review E, 83, 025102(R). https://doi.org/10.1103/PhysRevE.83.025102

Lambiotte, R., Rosvall, M., & Scholtes, I. (2019). From networks to optimal higher-order models of complex systems. Nature Physics, 15, 313–320. https://doi.org/10.1038/s41567-019-0459-y

Pan, R. K., & Saramäki, J. (2011). Path lengths, correlations, and centrality in temporal networks. Physical Review E, 84, 016105. https://doi.org/10.1103/PhysRevE.84.016105

Perra, N., Gonçalves, B., Pastor-Satorras, R., & Vespignani, A. (2012). Activity driven modeling of time varying networks. Scientific Reports, 2, 469. https://doi.org/10.1038/srep00469

Scholtes, I. (2017). When is a network a network? Multi-order graphical model selection in pathways and temporal networks. Proceedings of the 23rd ACM SIGKDD International Conference on Knowledge Discovery and Data Mining, 1037–1046. https://doi.org/10.1145/3097983.3098145

Zou, L., Ceria, A., & Wang, H. (2023). Short- and long-term temporal network prediction based on network memory. Applied Network Science, 8, 76. https://doi.org/10.1007/s41109-023-00597-w

## Supplementary results: network-specific primary coefficients

Table S1 gives all archived RQ1, RQ2, and RQ6 network-specific primary estimates used in the main text. BH-FDR corrections were applied post hoc within each RQ using only finite p-values with strictly positive finite standard errors (RQ1: 11; RQ2: 10; RQ6: 8). Nonestimable results are not counted as null effects; absent outputs are distinguished from nonestimable coefficient records.

*Table S1 (RQ1). Archived network-specific association coefficients, nominal 95% CIs and post hoc BH-FDR results.*

| Network | nobs | β | 95% CI | p | q (BH) |
|---|---|---|---|---|---|
| High School 2013 | 19,291 | +0.034237 | [+0.021121, +0.047354] | <0.0001 | <0.0001 |
| Workplace 2015 | 2,634 | +0.014074 | [-0.016609, +0.044757] | 0.3686 | 0.4055 |
| SFHH Conference | 7,656 | +0.046905 | [+0.030114, +0.063696] | <0.0001 | <0.0001 |
| Email-EU-core | 2,700 | -0.014852 | [-0.028153, -0.001550] | 0.0286 | 0.1050 |
| CollegeMsg | 3,134 | +0.004440 | [-0.004823, +0.013703] | 0.3475 | 0.4055 |
| Manufacturing E-mail | 7,935 | +0.006937 | [-0.000762, +0.014635] | 0.0774 | 0.1419 |
| Facebook Wall | 9,603 | -0.007461 | [-0.015196, +0.000275] | 0.0587 | 0.1292 |
| MathOverflow | 2,638 | +0.004525 | [-0.004749, +0.013799] | 0.3389 | 0.4055 |
| Bitcoin OTC | 3382 | NE | — | — | Nonestimable |
| NYC Yellow Taxi | 629,083 | -0.000561 | [-0.001592, +0.000470] | 0.2860 | 0.4055 |
| NYC Citi Bike | 1,277,810 | -0.000240 | [-0.000480, -0.000000] | 0.0495 | 0.1292 |
| Chicago Divvy | 185,963 | +0.000310 | [-0.000617, +0.001237] | 0.5117 | 0.5117 |

*Table S1 (RQ2). Archived network-specific association coefficients, nominal 95% CIs and post hoc BH-FDR results.*

| Network | nobs | β | 95% CI | p | q (BH) |
|---|---|---|---|---|---|
| High School 2013 | 9,664 | -0.036334 | [-0.082196, +0.009528] | 0.1205 | 0.4016 |
| Workplace 2015 | 1,422 | +0.023178 | [-0.100785, +0.147142] | 0.7140 | 0.9826 |
| SFHH Conference | 4,463 | -0.024666 | [-0.078994, +0.029663] | 0.3735 | 0.9339 |
| Email-EU-core | 2,006 | -0.002940 | [-0.101908, +0.096029] | 0.9536 | 0.9826 |
| CollegeMsg | 1,656 | +0.013680 | [-0.060973, +0.088334] | 0.7195 | 0.9826 |
| Manufacturing E-mail | 6,969 | -0.000309 | [-0.028055, +0.027438] | 0.9826 | 0.9826 |
| Facebook Wall | 2361 | NE | — | — | Nonestimable |
| MathOverflow | 912 | -0.004522 | [-0.079197, +0.070152] | 0.9055 | 0.9826 |
| Bitcoin OTC | 2326 | NE | — | — | Nonestimable |
| NYC Yellow Taxi | 586,496 | +0.008727 | [+0.004295, +0.013158] | 0.0001 | 0.0011 |
| NYC Citi Bike | 654,104 | +0.000924 | [+0.000109, +0.001739] | 0.0263 | 0.1317 |
| Chicago Divvy | 78,364 | +0.000678 | [-0.004085, +0.005442] | 0.7802 | 0.9826 |

*Table S1 (RQ6). Archived network-specific association coefficients, nominal 95% CIs and post hoc BH-FDR results.*

| Network | nobs | β | 95% CI | p | q (BH) |
|---|---|---|---|---|---|
| High School 2013 | 15,518 | +0.025259 | [+0.012377, +0.038140] | 0.0001 | 0.0005 |
| Workplace 2015 | 477 | +0.051989 | [+0.007596, +0.096383] | 0.0217 | 0.0434 |
| SFHH Conference | 5,553 | +0.005154 | [-0.007200, +0.017507] | 0.4135 | 0.4726 |
| Email-EU-core | — | — | — | — | No output |
| CollegeMsg | 803 | +0.005806 | [-0.018192, +0.029804] | 0.6354 | 0.6354 |
| Manufacturing E-mail | 4,608 | +0.013315 | [-0.001269, +0.027898] | 0.0735 | 0.1177 |
| Facebook Wall | — | — | — | — | No output |
| MathOverflow | — | — | — | — | No output |
| Bitcoin OTC | 1553 | NE | — | — | Nonestimable |
| NYC Yellow Taxi | 586,323 | -0.025129 | [-0.029582, -0.020676] | $<1\times10^{-12}$ | $<1\times10^{-12}$ |
| NYC Citi Bike | 1,108,476 | -0.000379 | [-0.000617, -0.000141] | 0.0018 | 0.0047 |
| Chicago Divvy | 129,156 | -0.000394 | [-0.000960, +0.000172] | 0.1722 | 0.2297 |